\documentclass[a4paper,11pt]{article}
\pdfoutput=1 

\usepackage{jheppub} 
\usepackage{subfigure}
\usepackage{xcolor}
\usepackage[T1]{fontenc} 
\usepackage{xspace}
\usepackage{multirow}
\usepackage{hyperref}
\usepackage{slashed}
\usepackage{amsmath}

\usepackage[utf8]{inputenc}

\newcommand{\Hone}{H\protect\scalebox{0.8}{1}\xspace}
\newcommand{\Zeus}{Z\protect\scalebox{0.8}{EUS}\xspace}
\newcommand{\HERA}{H\protect\scalebox{0.8}{ERA}\xspace}

\def\GeV{\ifmmode {\mathrm{\ Ge\kern -0.1em V}}\else \textrm{Ge\kern -0.1em V}\fi}%
\def\GeV{\ifmmode {\mathrm{\ Ge\kern -0.1em V}}\else \textrm{Ge\kern -0.1em V}\fi}%

\title{\boldmath Suppression of Electroweak Instanton Processes in High-energy Collisions}

\author{Valentin V. Khoze and}
\author{Daniel L. Milne}

\affiliation{IPPP, Department of Physics, Durham University, Durham DH1 3LE, UK}

\emailAdd{valya.khoze@durham.ac.uk}
\emailAdd{daniel.l.milne@durham.ac.uk}

\abstract{
Electroweak instantons are a prediction of the Standard Model and have been studied in great detail in the past although they have not been observed. Earlier calculations of the instanton production cross section at colliders revealed that it was exponentially suppressed at low energies, but may grow large at energies (much) above the sphaleron mass. Such calculations faced difficulty in the breakdown of the instanton perturbation theory in the high-energy regime. In this paper we review the calculation for the electroweak instanton cross section using the optical theorem, including quantum effects arising from interactions in the initial state and show that this leads to an exponential suppression of the cross section at all energies, rendering the process unobservable. 
}

\begin{document}
\preprint{IPPP/20/56}

\maketitle
\flushbottom


\section{\label{Sec:Intro}Introduction}
\medskip

Instantons are a prediction of the Standard Model, yet to be observed, arising from the non-trivial vacuum structure of non-Abelian gauge theories. 
Path integrals expanded around the instanton field configuration describe quantum tunnelling between different degenerate vacua in the Yang-Mills theory. 
In particle collisions at sufficiently high energies, instanton-dominated processes are characterised by high-multiplicity final states which are rare in normal SM backgrounds and which would lead to them being easy to separate from background in a detector environment. They have so far eluded detection as calculations in \cite{Ringwald:1989ee,Espinosa:1989qn,Mattis:1991bj,Rubakov:1996vz} predict the instanton cross-section to be exponentially suppressed at energies below the sphaleron mass scale,\footnote{The sphaleron mass is the height of the barrier that separates the adjacent topologically non-trivial vacua~\cite{Klinkhamer:1984di}. Its Standard Model value is
$M_{\rm sph} = f(m_H^2/m_W^2) \, \pi m_W/\alpha_W \simeq 9$ TeV. In this paper we will use the simpler parametrically similar to $M_{\rm sph}$ scale 
$E_{\rm np}$, defined in \eqref{eq:Enp}, which plays the role of the intrinsic non-perturbative scale of the weak sector of the Standard Model.}
\begin{equation}
E_{\rm np} \,= \,  \frac{\pi m_W}{\alpha_W}\,  \simeq\,  7.4\,{\rm TeV}.
\label{eq:Enp}
\end{equation}
It is only when partonic CoM energies become of the order of the sphaleron scale 
that one could expect electroweak instanton cross sections to become observable in a collider~\cite{Ringwald:1989ee,Espinosa:1989qn,Mattis:1991bj,Rubakov:1996vz,Ringwald:2003px}.

Instantons in QCD were recently investigated in \cite{Khoze:2019jta,Khoze:2020tpp} where it was shown that, at least in principle, QCD instantons are observable at the LHC. This is largely due to the fact that instantons produce a spherically symmetric final state and hence it is possible to discriminate from background using shape variables. This signal would be even more pronounced in the case of electroweak instantons due to the many final state leptons, coming from both the initial instanton process and the subsequent decay of gauge and Higgs bosons. The suppression of the instanton cross section is given by $e^{-2S_{I}}=e^{-\frac{4\pi}{\alpha}}$ and it is due to the fact that $\alpha_{s}\gg\alpha_{w}$ that QCD instantons are observable at the LHC while observation of electroweak instantons would require a collider such as the proposed FCC to reach sufficient CoM energies. 

Importantly, the function in the exponent also contains terms coming from the instanton--anti-instanton interaction and the effects of quantum corrections in the instanton background. The maximum cross section for the instanton process is then determined by the saddle point of this exponent. 
The calculation in  \cite{Khoze:2019jta,Khoze:2020tpp} critically relied on the inclusion of quantum effects due to interactions of the hard initial-state gluons in the instanton background, which was instrumental in cutting off contributions of QCD instantons of large size and thus resolving the infrared problem associated with them. It should be noted that electroweak instantons do not suffer from the infrared problem of large-size instantons in the same way as QCD instantons did, as the vev of the Higgs provides a cutoff on the instanton size. Nevertheless, the effect of the radiative exchanges between the highly energetic initial state partons, 
should lead in both cases to the suppression of the instanton rates in the asymptotic high-energy regime.\footnote{Most earlier studies of QCD instanton-induced processes were specific to deep-inelastic 
scattering (DIS)~\cite{Ringwald:1998ek,Ringwald:2003px}. In this case, it was the deep inelastic momentum scale $Q$ that
was essential for obtaining infrared safe instanton contributions in the DIS settings and at relatively low CoM energies.
The \Hone and \Zeus Collaborations have searched for QCD instantons at the \HERA collider~\cite{Adloff:2002ph, Chekanov:2003ww, H1:2016jnv}. }

In this paper we will, for the first time, take account a resummed quantum correction arising from exchange of weak gauge bosons in the initial state. We will show that this term leads to an exponential suppression of the cross section for the electroweak instanton at high energies, while the Higgs contributions suppress it in the low-energy range. Due to these effects, the saddle point value of the exponent is not greatly reduced compared to its maximum (negative) value and so the cross section remains suppressed and unobservable at all energies. 


\medskip
\section{\label{Sec:Calculation}Instanton rate}

The instanton is a solution of the classical equations of motion involving both the gauge fields and the fermions and additionally in the electroweak sector, the Higgs field. This solution tunnels between degenerate electroweak vacua which are labelled by integers known as the winding number or topological charge. The instanton is the solution of the equations of motion with topological charge, $Q=1$, while the anti-instanton has $Q=-1$. This solution was found in \cite{Belavin:1975fg} and many consequences were considered by t'Hooft in \cite{tHooft:1976snw,tHooft:1986ooh}. The instanton gauge field configuration in the singular gauge is:
\begin{equation}
A^{a}_{\mu}\left(x\right)=\,\frac{2}{g}\,\frac{\bar\eta_{a\mu\nu}\left(x-x_{0}\right)_{\nu}}{\left(x-x_0\right)^2 (\left(x-x_{0}\right)^{2}+\rho^{2})}\,,
\end{equation}
where $x_0$ is the instanton position, $\rho$ is the instanton size and the t'Hooft eta symbols, $\eta_{a\mu\nu}$, are defined in \cite{tHooft:1976snw}. 

We will consider the instanton-dominated electroweak process with two quarks in the initial state,
\begin{equation}
 q\, +\, q \, \to\, 7\,\bar{q} \,+\, 3\,\bar{l} + n_{w}\,W \, + n_{Z}\,Z \, +n_{H}\,H\,.
 \label{eq:inst1}
 \end{equation} 
The number of gauge and Higgs bosons in the final state is not fixed and may grow large. However the number of final state fermions is fixed at leading order by both the number of fermionic zero modes in the instanton background as well as by the anomalous violation of baryon and lepton number. It should be noted that at the energies considered in this paper all fermions are light and so we obtain one anti-fermion zero mode from each left handed doublet (with 3 for each quark due to colour) giving the above result. Note that two of the final state anti-fermion zero modes are inverted to become initial state fermions. 

\medskip
\subsection{Contributions of small instantons at large separations}
\medskip

The total cross-section $\sigma_{I}$ for instanton-generated $2\to anything$ processes
is most easily obtained from the imaginary part of the
forward elastic $2\to 2$ amplitude computed in the background of the instanton--anti-instanton ($I\bar{I}$) configuration,
\begin{eqnarray}
\sigma_{I} &=&  \frac{1}{E^2}\, {\rm Im} \, {\cal A}^{{I\bar{I}}}_{\,\,4} (p_1,p_2,-p_1,-p_2) 
\,,
\label{eq:op_th}
\end{eqnarray}
where $E=\sqrt{s} = \sqrt{(p_1+p_2)^2}$ is the CoM energy of the two incoming particles.
The amplitude itself is given by an integral over the instanton and anti-instanton parameters of a characteristic semi-classical exponential suppression factor, 
$e^{-\frac{4\pi}{\alpha_w} F}$,
that will be defined below\footnote{The normalisation by $\frac{4\pi}{\alpha_w}$ of the function $F$ in the exponent is chosen for convenience and to recall that at low energies the exponential suppression of the rate is in terms of twice the instanton action $S_I= \frac{2\pi}{\alpha_w}$.}  ({\it cf.}~\eqref{eq:calF1}), so that,
\begin{eqnarray}
\sigma_{I} \,\sim\,
{\rm Im} \int d\rho\,
d\bar{\rho} \, d^4 R \, d^3 u  \,\, e^{-\frac{4\pi}{\alpha_w} F}\,.
 \label{eq:sig1}
\end{eqnarray}
The integration variables in \eqref{eq:sig1} are the  instanton and anti-instanton sizes  $\rho$ and $\bar{\rho}$, the separation vector between the $I$ and $\bar{I}$  positions $R_\mu=(R_0,\vec{R})$, and the unit 4-vector $u_\mu$ that characterises the relative orientation between $I$ and $\bar{I}$ in the $SU(2)$ space.
At present we are primarily interested in the exponent in  \eqref{eq:sig1} and thus we have ignored for now all non-exponential terms in the integrand.
The function in the exponent reads,
\begin{eqnarray}
-\,\frac{4\pi}{\alpha_w} F\,\,=\,
-\,S_{I\bar{I}}\,+\,
E R_0
\,-\, \frac{\alpha_w}{16\pi}(\rho^2+\bar{\rho}^2) \, E^2\, \log \left({E^2}/{m_W^2}\right)\,,
 \label{eq:calF1}
\end{eqnarray}
where the first factor is the Euclidean action of the instanton--anti-instanton configuration, the second term, $E R_0$, comes from the initial and final particles in the forward elastic scattering amplitude~\cite{Zakharov:1990dj,Khoze:1990bm} and the third term is the quantum effect due to resummed hard radiative corrections in the instanton and anti-instanton background~\cite{Mueller:1990ed}.

At large instanton--anti-instanton separation, $R/\rho \gg 1$, (anti-)instantons of small sizes, $\rho \, m_W  \ll 1$, interact weakly and the 
$I\bar{I}$ action in the electroweak theory can be computed in the nearly dilute instanton gas limit, giving a
 well-known expression (see e.g.~\cite{Zakharov:1990dj,Khoze:1990bm}),
\begin{equation}
S_{I\bar{I}}\,=\, \frac{4\pi}{\alpha_w}\left(1\,-\, 2\frac{\rho^2\bar{\rho}^2}{R^4}\Bigl(\frac{4(u\cdot R)^2}{R^2}-1\Bigr)
\,+\, \frac{1}{4}\left( (m_W \rho)^2 +(m_W\bar{\rho})^2 \right)
\right) +\,  \ldots\,.
 \label{eq:SII1}
\end{equation}
The first term $\propto 1$ in brackets in \eqref{eq:SII1} represents twice the classical instanton action, the second term $\propto (\rho/R)^4$ is the leading-order interaction potential between the instanton and the anti-instanton (it is of the dipole-dipole interaction form in 4D), and the final term $\propto (m_W \rho)^2$ is the Higgs field contribution to the instanton--anti-instanton action.\footnote{The dots in \eqref{eq:SII1} 
stand for the omitted higher order terms in the large-separation and small instanton size expansion,
 $\frac{4\pi}{\alpha_w} \left({\cal O}(\rho/R)^6 +{\cal O}((\rho/R)^2(m_W\rho)^2)+ {\cal O}(m_W\rho)^4\right)$.}

It then follows that in the weak coupling limit  $\frac{4\pi}{\alpha_w} \gg 1$, the 
integral in \eqref{eq:sig1} is dominated by the saddle-point of ${\cal F}$ in \eqref{eq:calF1}. The saddle-point is given by,
\begin{equation}
u_\mu = R_\mu/R\,,\quad R_\mu =( R_{0*}, \vec{0})\,,\quad  \rho=\bar{\rho}=\rho_*\,,
\label{eq:sp1}
\end{equation}
where
\begin{equation}
R_{0*} =\,\frac{6^{1/3}}{m_W}\,\frac{\varepsilon^{1/3}}{\left(1+\frac{\varepsilon^2}{8} \log\left( \frac{\varepsilon\, \pi}{\alpha_w}\right)\right)^{2/3}}\,\,,
\quad
\rho_*=\, \frac{6^{1/6}}{2 \,m_W} \,\frac{\varepsilon^{2/3}}{\left(1+\frac{\varepsilon^2}{8} \log\left(\frac{\varepsilon\, \pi}{\alpha_w}\right)\right)^{5/6}}\,.
\label{eq:sp2}
\end{equation}
In the equations above we introduced the parameter $\varepsilon$ as the ratio of the collision energy $E$ to the sphaleron-induced non-perturbative 
scale $E_{\rm np}$,
\begin{equation}
\varepsilon \,=\, \frac{E}{\pi m_W/\alpha_w} \,=\, \frac{E}{E_{\rm np}}\,.
\label{eq:varpsdef1}
\end{equation}

\medskip

Now let us consider the saddle-point value of the function $\cal F$ in \eqref{eq:calF1}
that appears in the exponent of the semi-classical instanton rate, computed in the low-energy and the high-energy regimes.
To this end we would like to rewrite the saddle-point solution \eqref{eq:sp2} in terms of the dimensionless variables 
$\chi = R_{0*}/\rho_*$ and $\hat{\rho}= \rho_* m_W$ representing the $I\bar{I}$ separation and the scale size. 
We find,
\begin{eqnarray}
\chi \,=\, 6^{1/6}  2 \,\, \frac{\left(1+\frac{\varepsilon^2}{8} \log\left(\frac{\varepsilon\, \pi}{\alpha_w}\right)\right)^{1/6}}{\varepsilon^{1/3}}
&\to&
\left\{
                \begin{array}{ll}
                  6^{1/6} \, 2 \cdot \varepsilon^{-1/3}\,, \quad \varepsilon \to 0\\
                 2^{2/3} 3^{1/6}\cdot \left( \log\left(\frac{\varepsilon\, \pi}{\alpha_w}\right)\right)^{1/6}, \quad \varepsilon \to \infty
                \end{array}
              \right.
\label{eq:chi1}\\
\nonumber\\
\hat{\rho} \,=\, \frac{6^{1/6}\, \varepsilon^{2/3}}{2 \left(1+\frac{\varepsilon^2}{8} \log\left(\frac{\varepsilon\, \pi}{\alpha_w}\right)\right)^{5/6}}
&\to&
\left\{
                \begin{array}{ll}
                  6^{1/6}  \, 2^{-1} \cdot \varepsilon^{2/3}\,, \quad \varepsilon \to 0\\
                 6^{1/6}\,2\sqrt{2} \cdot \varepsilon^{-1} \left( \log\left(\frac{\varepsilon\, \pi}{\alpha_w}\right)\right)^{-5/6}, \,\, \varepsilon \to \infty
                \end{array}
              \right.
\label{eq:rhohat1}
\end{eqnarray}
From \eqref{eq:rhohat1} it follows that the the variable $\hat{\rho}$ that characterises the (anti-)instanton size is small 
(i.e. power-like suppressed with $\varepsilon$) in both, the low-energy and the high-energy regimes. Thus it is justifiable to neglect the 
higher-order-in-$\hat{\rho}$ corrections to the expression \eqref{eq:calF1} for $\cal F$.

The inverse separation $1/\chi$ is small in the low-energy regime, $1/\chi \sim \varepsilon^{1/3} \to 0$, while in the opposite regime of asymptotically high energies, it is only logarithmically suppressed,  $1/\chi \sim \log^{-1/6} (E/m_W)$. This implies that we need to take into account higher order corrections 
in powers of $1/\chi$ to the semiclassical exponent in \eqref{eq:calF1}. Hence we conclude from Eqs.~\eqref{eq:chi1}-\eqref{eq:rhohat1} that we can continue working with small instantons, but in the high-energy limit will also have to include higher-order-in-$\rho/R$ corrections to the dipole-dipole interactions of instantons. 
This will be done in the following section.

For now, we can present the saddle-point value of the semiclassical exponent, based on the leading-order expression.
It is instructive to cancel the factor of $4\pi/\alpha_w$ on both sides of Eq.~\eqref{eq:calF1}
and define the function $S(\chi)$ via,
\begin{eqnarray}
S_{I\bar{I}} \,=\,  \frac{4\pi}{\alpha_w}\, \left( S(\chi) \,+\, \frac{1}{2} \hat{\rho}^2\right)\,,
\label{eq:Schidef}
\end{eqnarray}
where $S(\chi)$ corresponds to the pure gauge field part of the $I\bar{I}$ action  \eqref{eq:SII1}. 
In the maximally attractive relative orientation channel, $u_\mu = R_\mu/R$, we have,
\begin{equation}
S(\chi)\,=\, 1\,-\, 6/\chi^4 +\,  \ldots\,.
 \label{eq:SII2}
\end{equation}

Using 
the saddle-point equations, we find a surprisingly compact expression for the holy-grail function $F$ computed on the saddle-point,\footnote{We note that 
$F_*$ in \eqref{eq:Fsp1} depends only on the $\chi$ variable, while the dependence on $\hat{\rho}$ has cancelled out.}
\begin{eqnarray}
F_*\,=\, S(\chi)\,-\,\frac{1}{2}\, \chi\, S'(\chi)\,.
\label{eq:Fsp1}
\end{eqnarray}
At energies much below the sphaleron mass we reproduce the well-known low-energy expression~\cite{Zakharov:1990dj,Khlebnikov:1990ue,Khoze:1990bm}, 
\begin{eqnarray}
F_*&=& 1\,-\, \frac{18}{\chi^4} \, +\,  {\cal O}(1/\chi^6) \,=\, 1 \,-\, \frac{6^{1/3}3}{2^4} \, \varepsilon^{4/3} \,+\,{ \cal O}(\varepsilon^2)\,,
\label{eq:Fsp111} \\
\sigma_I &\sim& e^{-\frac{4\pi}{\alpha_w}\left(1-\frac{6^{1/3}3}{2^4} \, \varepsilon^{4/3}\right)} \,\ll 1\, \qquad {\rm for}\,\,
\varepsilon \ll 1\,.
\nonumber
\end{eqnarray}
To access the high-energy regime, we need to extend this analysis to finite separations. 

\medskip
\subsection{Accounting for the higher order terms in $I\bar{I}$ interactions}
\medskip

The action of the instanton--anti-instanton configuration is known in the electroweak theory to next-to-next-to leading order 
in the expansion in inverse separation, $\rho/R$, and the instanton scale size, $m_w \rho$, parameters. The expression generalising the leading-order formula
\eqref{eq:SII1} was derived in \cite{Balitsky:1993xc}. In terms of the dimensionless variables,
\begin{equation}
\chi \,=\, R/\rho\,, \qquad \hat{\rho}\,=\, m_W \rho\,,
\end{equation}
for the symmetric configuration $\rho=\bar{\rho}$ in the maximally attractive $I\bar{I}$ interaction 
channel, the instanton--anti-instanton action reads,
\begin{eqnarray}
S_{I\bar{I}} \,=&&  \frac{4\pi}{\alpha_w}\, \left[ S(\chi) \,+\, \frac{1}{2} \,\hat{\rho}^2 \left(1+\frac{1}{\chi^2}-\frac{12}{\chi^4}\log \chi\right)
\right.
\label{eq:BSaction}
\\
&& \left. \quad +\, \frac{1}{2} \,\hat{\rho}^4 \left(\Bigl(\frac{3}{2}-\frac{1}{2}\frac{m_H^2}{m_W^2}\Bigr)\log \chi
+ \Bigl(\frac{3}{2}-\frac{m_H^2}{m_W^2}\Bigr) \log \hat{\rho}
- \frac{m_H^2}{m_W^2} \log\Bigl(\frac{m_H}{m_W}\Bigr)
\right)
\right]\,.
\nonumber
\end{eqnarray}
As before, $S(\chi)$ appearing on the right hand side of \eqref{eq:BSaction}, 
denotes the terms in the normalised $I\bar{I}$ action that arise solely from the gauge field components of the instanton--anti-instanton configuration. 
To the order
${\cal O}(\chi^{-8} \log \chi)$ it reads,
\begin{equation}
S(\chi) \,=\, 1\,-\, \frac{6}{\chi^4} \,+\, \frac{24}{\chi^6} \,+\, \frac{72}{\chi^8} \, \log \chi \,.
 \label{eq:SII3}
\end{equation}
The remaining terms in the square brackets in \eqref{eq:BSaction} account for the Higgs field components and their interactions with the gauge fields, including the mass effects.
The overall expression \eqref{eq:BSaction} gives  the $I\bar{I}$ action up to the order $\hat{\rho}^4$, 
$\hat{\rho}^2 \chi^{-4}$ and $\chi^{-8}$.
The corresponding instanton holy-grail function $F$  \eqref{eq:calF1} then takes the form,
\begin{eqnarray}
\label{eq:fullF}
F \,=&&
-\,\frac{1}{4}\, \varepsilon \hat{\rho}\chi\,+\, S(\chi) \,+\,  \frac{1}{2} \,\hat{\rho}^2 \left(1+\frac{\varepsilon^2}{8} \, \log\left(\frac{\varepsilon\, \pi}{\alpha_w}\right)
+\frac{1}{\chi^2}-\frac{12}{\chi^4}\log \chi\right)
 \label{eq:calF11} \\
&& \left. \quad +\, \frac{1}{2} \,\hat{\rho}^4 \left(\Bigl(\frac{3}{2}-\frac{1}{2}\frac{m_H^2}{m_W^2}\Bigr)\log \chi
- \Bigl(\frac{3}{2}-\frac{m_H^2}{m_W^2}\Bigr) \log \hat{\rho}
- \frac{m_H^2}{m_W^2} \log\Bigl(\frac{m_H}{m_W}\Bigr)
\right)
\right].
\nonumber
\end{eqnarray}
The saddle-point solution, $\hat{\rho}(\varepsilon)$, $\chi(\varepsilon)$, is determined by solving the equations,
\begin{equation}
\partial_\chi F(\chi,\hat{\rho}; \varepsilon) \,=\, 0\, \qquad \partial_{\hat{\rho}} \,F (\chi,\hat{\rho}; \varepsilon)  \,=\, 0\,,
\label{eq:speqs}
\end{equation}
which is then used to compute the value of $F$ on the saddle-point, $F_*(\varepsilon)$, that dominates the instanton cross-section integral 
\eqref{eq:sig1} in the steepest-descent a.k.a. the semi-classical approximation.

\medskip
\subsubsection{The low-energy limit}
\medskip

At energies much below the sphaleron scale, $\varepsilon \ll 1$, we can neglect the $\frac{\varepsilon^2}{8} \, \log\left(\frac{\varepsilon\, \pi}{\alpha_w}\right)$ term on the first line in
\eqref{eq:calF11} relative to 1, and from this determine the value of $F_*(\varepsilon)$  at the corresponding saddle-point solution:
\begin{eqnarray}
F_* \,=\, 1 \,-\, \frac{6^{1/3}3}{2^4} \, \varepsilon^{4/3} \,+\, \frac{3}{32}\, \varepsilon^2
+\, \frac{1}{64}\,\frac{3^{2/3}}{2^{1/3}} \left(\frac{m_H^2}{m_W^2}-\frac{4}{3}\right) \varepsilon^{8/3} \log(1/{\varepsilon})
\,+\,{ \cal O}(\varepsilon^{8/3})\,.
\label{eq:Fsp2} 
\end{eqnarray}
The first three terms on the right hand side of \eqref{eq:Fsp2} were computed in \cite{Khoze:1990bm} and
the NNLO term in the $\varepsilon$ expansion $\sim \varepsilon^{8/3} \log(1/{\varepsilon})$ was derived in \cite{Balitsky:1993xc}. 

Our main objective is to go beyond the known low-energy regime established by \eqref{eq:Fsp2}.
To this end we will now consider the opposite extreme of the high-energy limit, $\varepsilon \gg1$ with the ultimate aim to be able to match the two regimes at intermediate values of $\varepsilon$ where the energy is of the order of the sphaleron mass.

\medskip
\subsubsection{The high-energy limit}
\medskip

As we already noted in the previous section, instantons remain small in the high-energy range;
according to the second equation in \eqref{eq:rhohat1} the dominant value of the instanton size is
$\hat{\rho}\, \sim \varepsilon^{-1} \log^{-5/6}(\varepsilon \pi/\alpha_w)  \to 0$ when $\varepsilon \to \infty$.

Thus we can neglect all $\sim \hat{\rho}^4$ terms and most of the $\sim \hat{\rho}^2$ terms in the expression for $F$ on the right hand side of \eqref{eq:calF11}.
The only $\sim \hat{\rho}^2$ term that we must retain in the $\varepsilon \to \infty$ limit is the term enhanced by $\varepsilon^2$, namely
the $\frac{1}{2} \,\hat{\rho}^2\,\frac{\varepsilon^2}{8} \, \log \frac{\varepsilon\, \pi}{\alpha_w}$ contribution arising from the resummed hard quantum corrections in the $I\bar{I}$ background.
With this we have,
\begin{eqnarray}
F \,=&&
-\,\frac{1}{4}\, \varepsilon \hat{\rho}\chi\,+\, S(\chi) \,+\,  \frac{1}{2} \,\hat{\rho}^2 \left(\frac{\varepsilon^2}{8} \, \log\Bigl(\frac{\varepsilon\, \pi}{\alpha_w}\Bigr)
+1
\right).
 \label{eq:calF22} 
 \end{eqnarray}
In practice, to facilitate the comparison between the 
high-energy and the low-energy limits, we have also retained the order-1 term in addition to the order-$\varepsilon^2$ term in the brackets 
on the right hand side of \eqref{eq:calF22}.

Our earlier analysis of the saddle-point solution for the $I\bar{I}$ separation variable $\chi$ in the second equation in
\eqref{eq:chi1}, indicates that $1/\chi$ is only logarithmically suppressed in the high-energy limit, and hence we cannot continue using the 
large separation approximation to the
gauge-field instanton--anti-instanton action, $S(\chi)$, in the form \eqref{eq:SII3}.
We need the expression for $S(\chi)$ valid at all finite separations. 

This is where the gradient flow approach, a.k.a the valley method, of Balitsky and Yung~\cite{Balitsky:1986qn,Yung:1987zp} for constructing instanton--anti-instanton configurations at arbitrary values of instanton and anti-instanton collective coordinates in gauge theory 
becomes indispensable and solves the problem.
Using the conformal invariance of classical Yang-Mills theory, in Ref.~\cite{Yung:1987zp} 
Yung found an exact solution of the valley equation for the instanton--anti-instanton gauge field
configuration, $A_\mu^{I\bar{I}}$. The action on this configuration was computed in \cite{Khoze:1991sa,Khoze:1991mx,Verbaarschot:1991sq} and gives,
\begin{eqnarray}
S^{\rm \, gauge}_{I\bar{I}}(z) \,=\, \frac{16\pi^2}{g^2} \left(3\frac{6z^2-14}{(z-1/z)^2}\,-\, 17\,-\, 
3 \log(z) \left( \frac{(z-5/z)(z+1/z)^2}{(z-1/z)^3}-1\right)\right)
 \label{eq:Sgauge}
\end{eqnarray}
where  the variable $z$ is the conformal ratio of the (anti-)instanton collective coordinates,
\begin{equation}
z\,=\, \frac{R^2+\rho^2+\bar{\rho}^2+\sqrt{(R^2+\rho^2+\bar{\rho}^2)^2-4\rho^2\bar{\rho}^2}}{2\rho\bar{\rho}}
\,,
\label{eq:defz}
\end{equation}
which plays the role of the single negative mode of the instanton--anti-instanton valley configuration along which the instantons attract each other.

\begin{figure}[]
\begin{center}
\begin{tabular}{cc}
\hspace{-.4cm}
\includegraphics[width=0.5\textwidth]{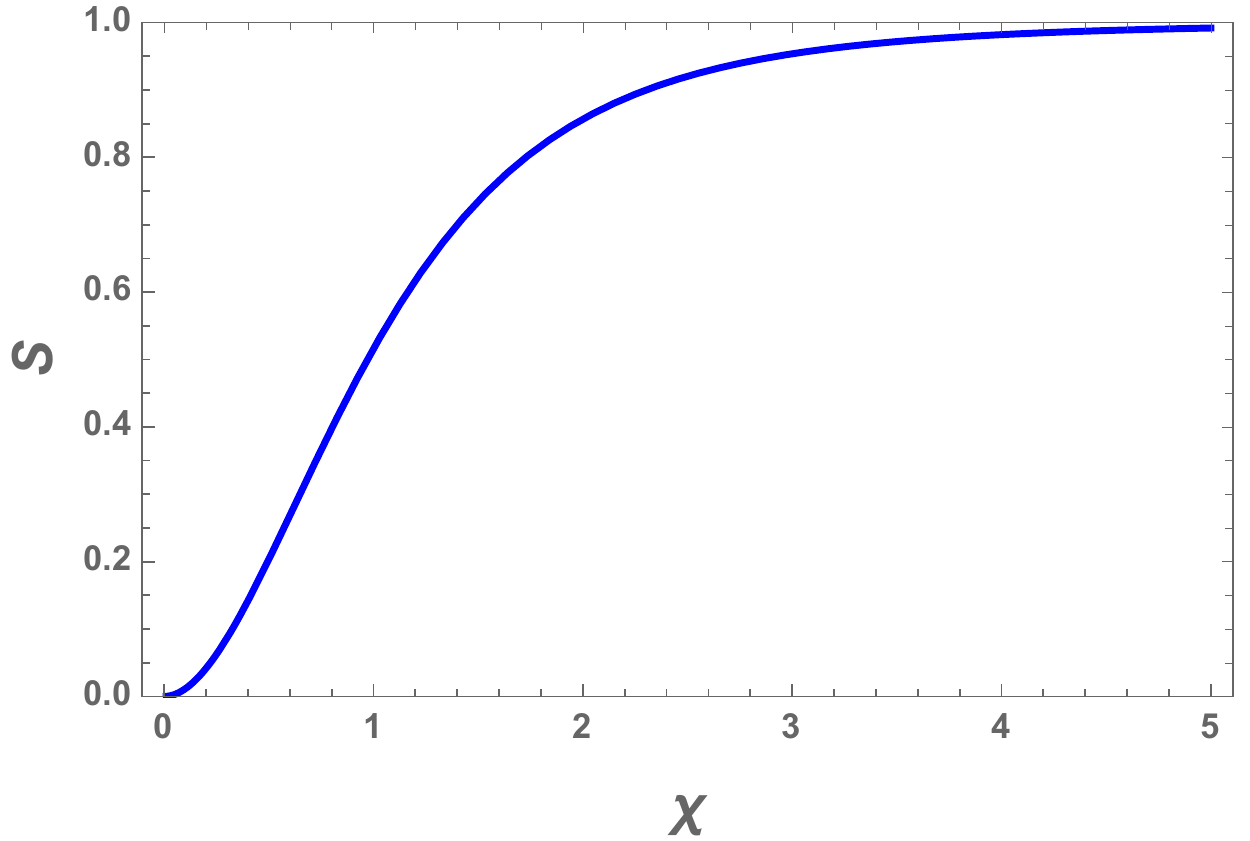}
&
\includegraphics[width=0.5\textwidth]{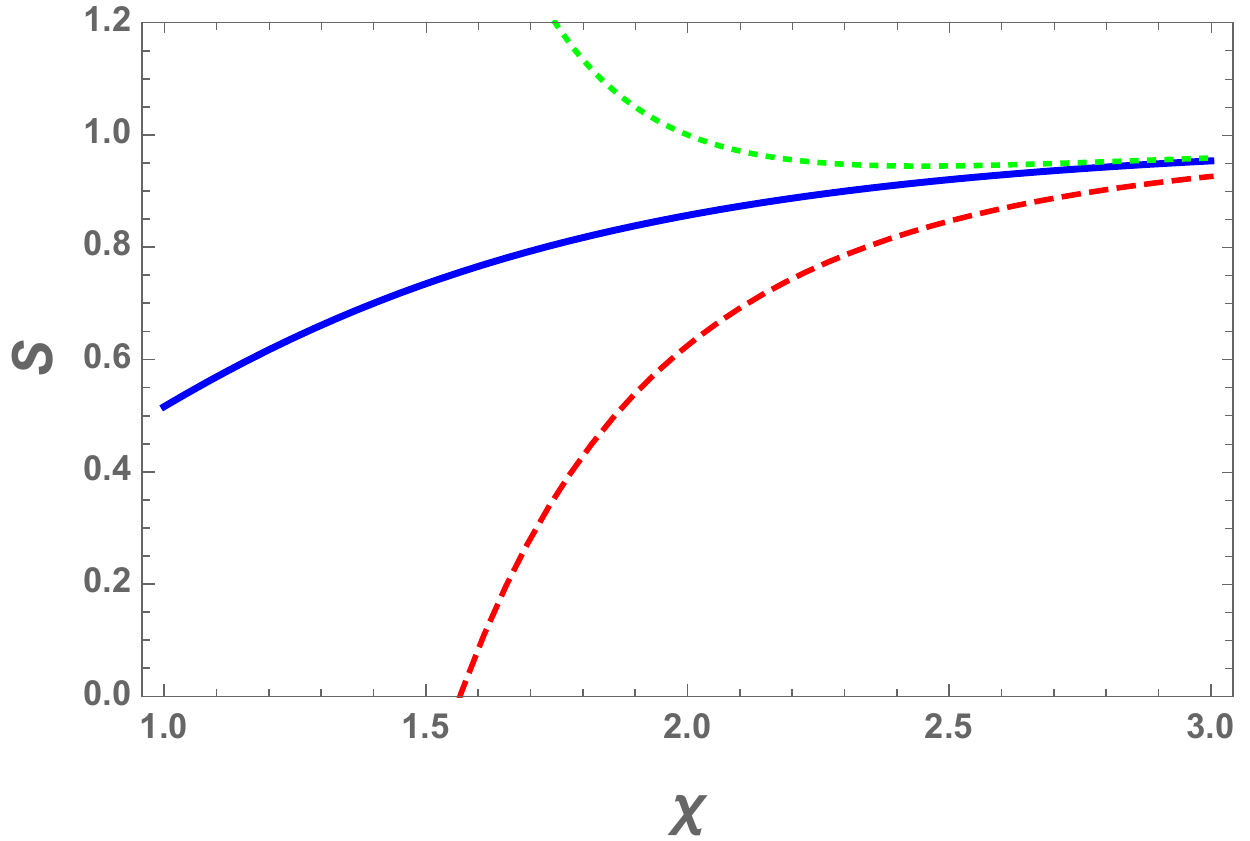}
\\
\end{tabular}
\end{center}
\vskip-.4cm
\caption{
The action \eqref{eq:Sz11} of the instanton--anti-instanton gauge configuration as a function of $\chi=R/\rho$ (solid line).
$S(\chi)$ approaches one
as $\chi\to \infty$ where the interaction potential vanishes, and $S\to 0$ as $\chi\to 0$ where the
 instanton and the anti-instanton 
annihilate into the perturbative vacuum.
The plot on the right also shows
the leading-order (dashed line) and the next-to-leading-order (dotted line) approximations in \eqref{eq:SII3}. The regime
of interest to us for matching the high-energy and the low-energy regimes is $\chi \sim 2$, which is far away from the perturbative annihilation region.}
\label{fig:S}
\end{figure}

The expression for $S(\chi)$ that we need to use on the right hand side of \eqref{eq:calF22} follows from the above,
\begin{eqnarray}
S(\chi) &=& 3\frac{6z^2-14}{(z-1/z)^2}\,-\, 17\,-\, 
3 \log(z) \left( \frac{(z-5/z)(z+1/z)^2}{(z-1/z)^3}-1\right)\biggl|_{z=z(\chi)}
 \label{eq:Sz11}\\
z(\chi) &=& \frac{1}{2} \left(\chi^2 +\chi\sqrt{\chi^2+4}+2\right)\,.
 \label{eq:zchi11}
\end{eqnarray}
${\cal S}(\chi)$ is plotted in Fig.~\ref{fig:S}.

\begin{figure}[]
\begin{center}
\begin{tabular}{cc}
\hspace{-.4cm}
\includegraphics[width=0.5\textwidth]{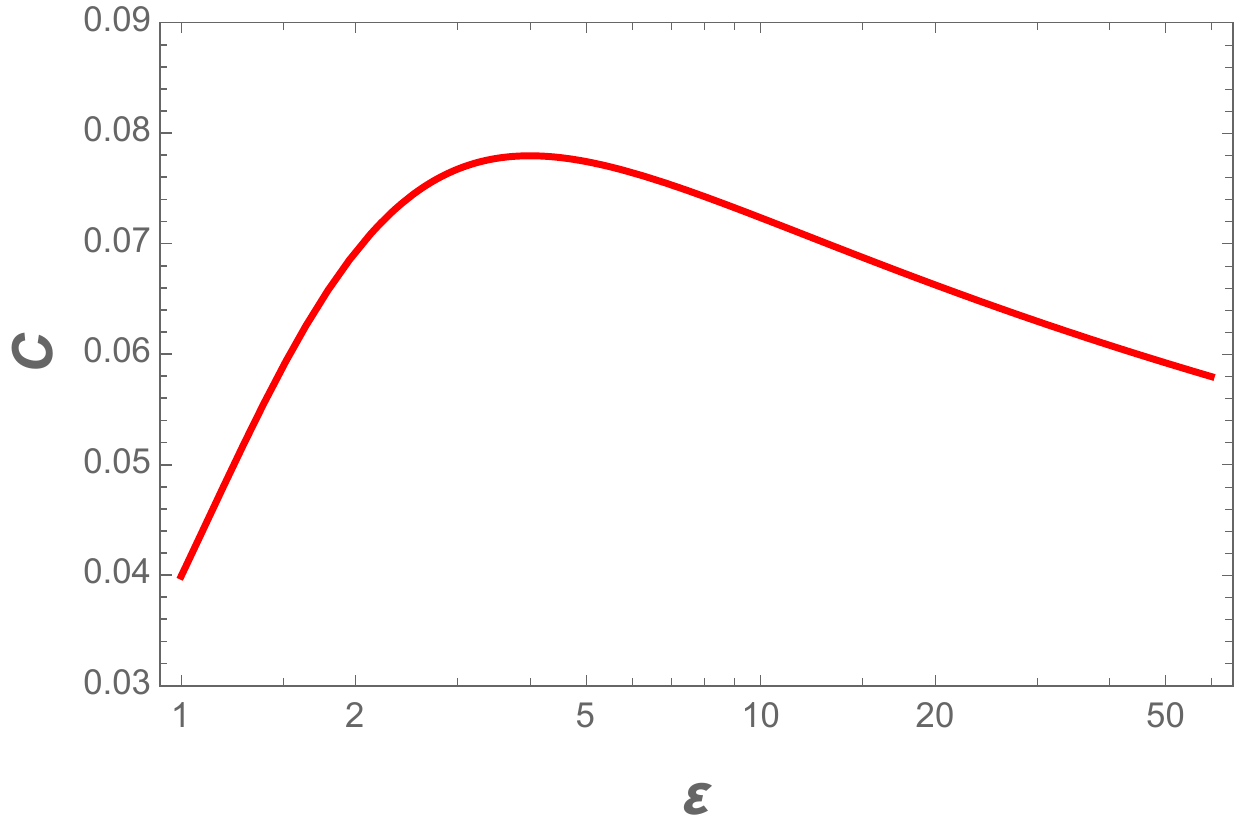}
&
\includegraphics[width=0.5\textwidth]{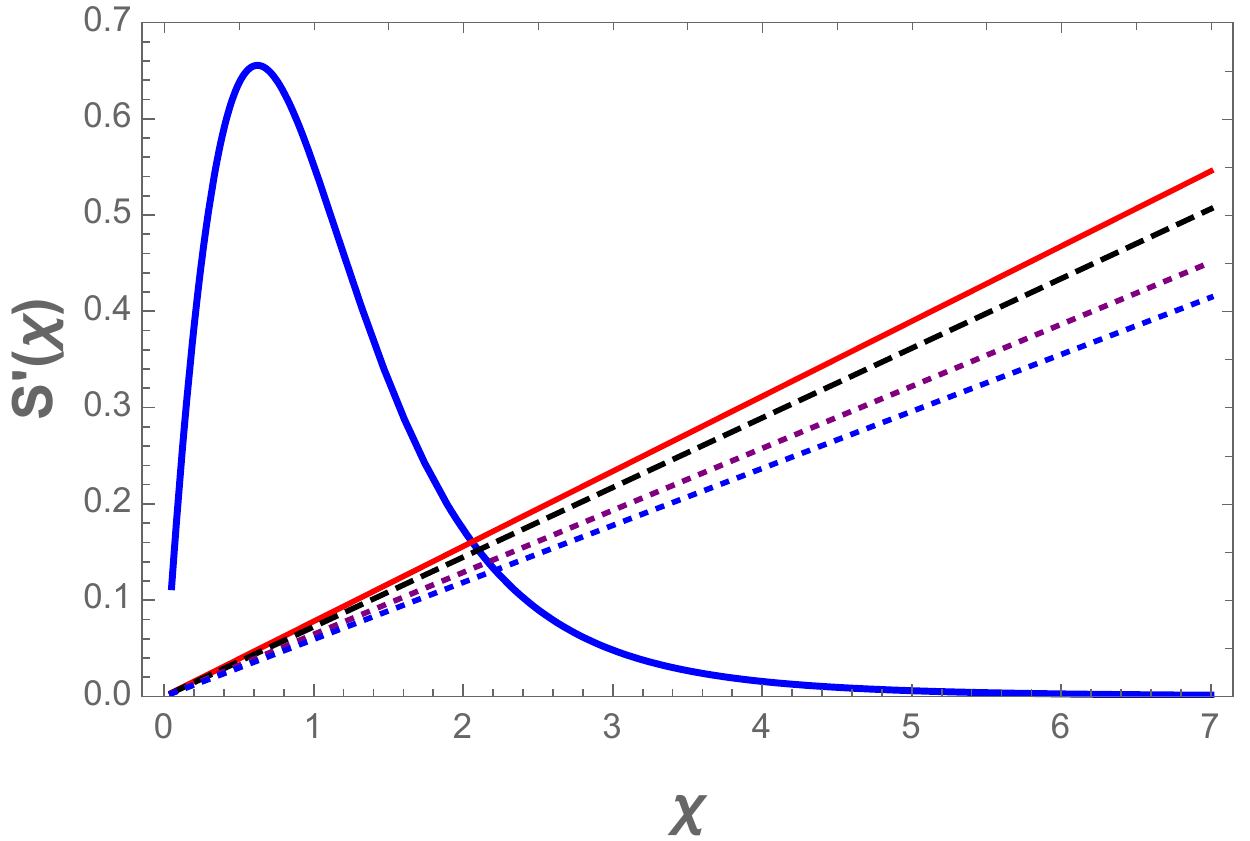}
\\
\end{tabular}
\end{center}
\vskip-.4cm
\caption{
The plot on the left depicts the function $C(\varepsilon)$ in \eqref{eq:coefC} with a local maximum at $\varepsilon =4$. The plot on the right shows the derivative of the action $S'(\chi)$ 
(solid blue line) and  the linear function $C(\varepsilon) \chi$ for various values of the energy $\varepsilon$. The steepest possible slope of $C(\varepsilon) \chi$ 
is at $\varepsilon =4$ (solid red line), with higher energies represented by $\varepsilon =10$ (dashed black), $\varepsilon =25$ (dotted purple)
and $\varepsilon =50$ (dotted blue line).
The saddle-point values of $\chi$ are given by the intersection points of the $S'(\chi)$ curve with the lines $C(\varepsilon) \chi$ for each value of $\varepsilon$.}
\label{fig:Sprime}
\end{figure}

The saddle-point equations \eqref{eq:speqs} for $F$ given in \eqref{eq:calF22} read,
\begin{equation}
\frac{1}{4}\, \varepsilon \hat{\rho}\,=\, S^\prime (\chi)\,,\quad
\frac{1}{4}\, \varepsilon \chi\,=\, \hat{\rho} \left(\frac{\varepsilon^2}{8} \, \log\Bigl(\frac{\varepsilon\, \pi}{\alpha_w}\Bigr)+1
\right).
\label{eq:speqs2} 
\end{equation}
Their solutions are given by:
\begin{eqnarray}
\frac{\chi}{2 \left(
\log\left(\frac{\varepsilon \pi}{\alpha_w}\right) 
+{8}\,{\varepsilon^{-2}}\right)} &=& S^\prime (\chi)\,,
\label{eq:solnchiV}  \\
\hat{\rho} &=& {\varepsilon^{-1}}\, 4  S^\prime (\chi)\,,
\label{eq:solrhoV}
\end{eqnarray}
where \eqref{eq:solnchiV} gives the equation for $\chi$ that can be solved numerically or graphically, as indicated in Fig.~\ref{fig:Sprime}.
The solution for $\hat{\rho}$ is then read off the second equation \eqref{eq:solrhoV}.

The equation \eqref{eq:solnchiV} is satisfied when the
coefficient in front of $\chi$, 
\begin{equation}
C(\varepsilon)\,=\, \frac{1}{2 \left(
\log\left(\frac{\varepsilon \pi}{\alpha_w}\right) 
+\frac{8}{\varepsilon^2}\right)}\, ,
\label{eq:coefC}
\end{equation}
matches the slope of $S(\chi)$. This coefficient is plotted on the left panel in Fig.~\ref{fig:Sprime}
and has a maximum at $\varepsilon =4$,
\begin{equation}
{\rm max}_\varepsilon\,  C(\varepsilon) \,=\, 
\frac{1}{2 \log\left(\frac{4 \pi}{\alpha_w}\right) +1}\,\simeq\, 0.078\,, \quad {\rm at} \,\,\, \varepsilon=4\,.
\end{equation}
The $I\bar{I}$ separation $\chi$ is minimal, $\chi_{\rm min}=2.06$, at the critical value of $\varepsilon=4$ and grows at both higher, $\varepsilon >4$, and lower, $\varepsilon < 4$, energies.

The holy-grail function \eqref{eq:calF22} computed on the solutions $\chi(\varepsilon)$, $\hat{\rho}(\varepsilon)$ at the saddle-point 
 \eqref{eq:solnchiV}-\eqref{eq:solrhoV} takes the form,
 \begin{eqnarray}
F_*\,=\, S(\chi)\,-\,\frac{1}{2}\, \chi\, S'(\chi)\,,
\label{eq:Fspfin}
\end{eqnarray}
 where we made use of the saddle-point equations 
\eqref{eq:speqs2} in arriving at the compact expression above.
Note that $F_*$ is given by the same formula in terms of $S$ as in the low-energy limit in \eqref{eq:Fsp1}.
  
Table~\ref{Tab:one} presents a selection of the saddle-point values $\chi(\varepsilon)$, $\hat{\rho}(\varepsilon)$ along with the corresponding values for the
normalised $I\bar{I}$ action $S(\chi)$ and the holy grail function $F_*$ 
in the high-energy range, starting with  $\varepsilon=30$ and going down in energy past the turning point $\varepsilon=4$, where $F_*$ is minimal, down to a minimum energy of $\epsilon=0.5$.
\begin{table}[]
\begin{center}
\begin{tabular}{|r||r|r|r||r|}
\hline
$\varepsilon$ & $\chi\,\,$ & $\hat{\rho}\,\,\,$ & $S(\chi)$ & $F_*\,\,$\\
\hline\hline
30   &   2.178 & 0.018 & 0.884 & $0.735$\\ 
20   &   2.150 & 0.028 & 0.880 & $0.727$\\ 
10   &   2.101 & 0.061& 0.873 & $0.713$\\ 
7.0   &   2.078 & 0.089 & 0.870 & $0.707$\\
5.0   &   2.063 & 0.128 & 0.867 & $0.702$\\ 
\hline
4.0   &   2.059 & 0.160 & 0.867 & $0.701$\\ 
\hline
3.0   &   2.068 & 0.212& 0.868 & $0.704$\\ 
2.0   &   2.125 & 0.294& 0.877 & $0.720$\\ 
0.5   &   3.090 & 0.345& 0.958 & $0.891$\\ 
\hline
\end{tabular}
\end{center}
\vskip-.4cm
\caption{Saddle-point solution
for the instanton separation, $\chi=R/\rho$ and the instanton size $\hat{\rho}$, along with the values for
the $I\bar{I}$ action $S(\chi)$ and the instanton suppression function $F_*$ in the range from $0.5\le \varepsilon=E/(\pi m_W/\alpha_w)\le 30$. The saddle-point was obtained using the simplified analytic expression for $F$ in \eqref{eq:calF22}.
Instantons are exponentially suppressed at all energies with the minimal value of $F_*$ at the critical energy $\varepsilon=4$. }
\label{Tab:one}
\end{table}

 \begin{figure}[]
\begin{center}
\begin{tabular}{cc}
\hspace{-.4cm}
\includegraphics[width=0.5\textwidth]{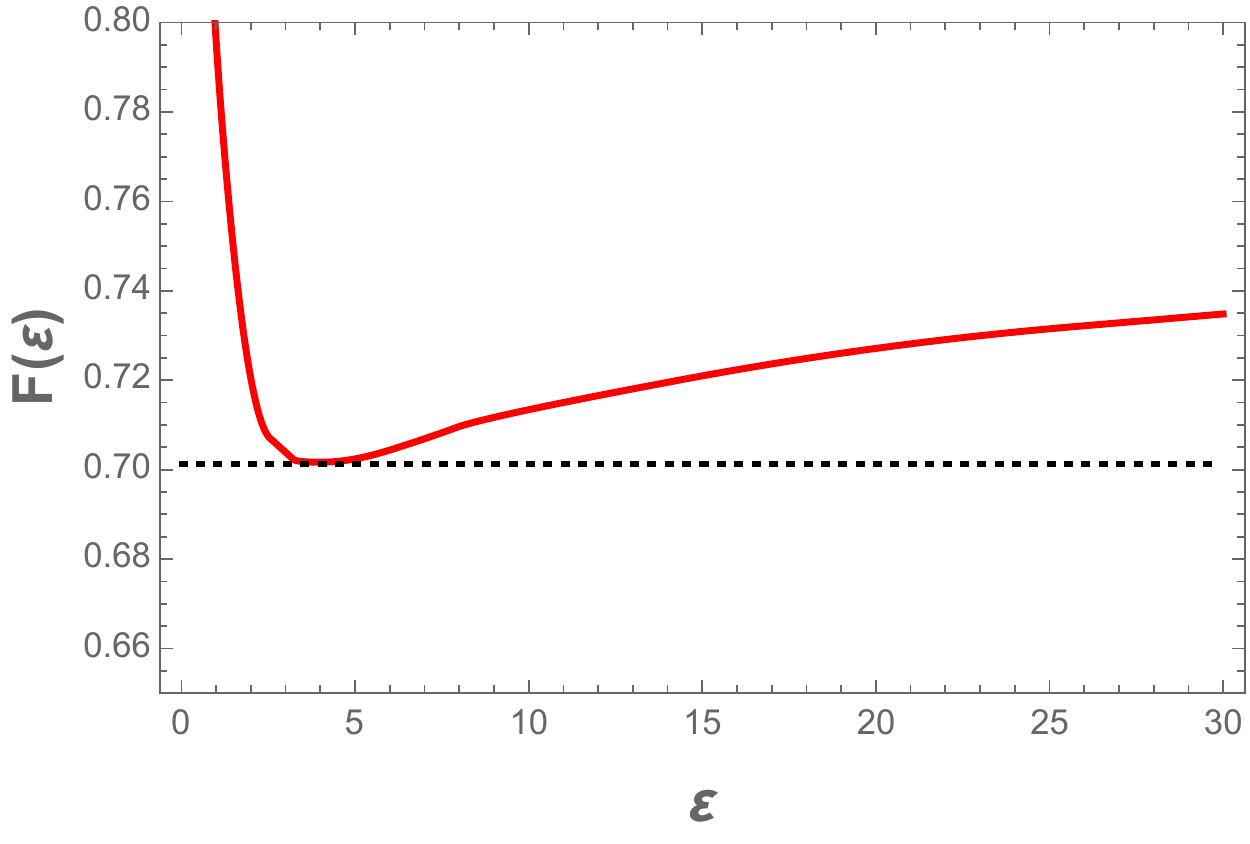}
&
\includegraphics[width=0.5\textwidth]{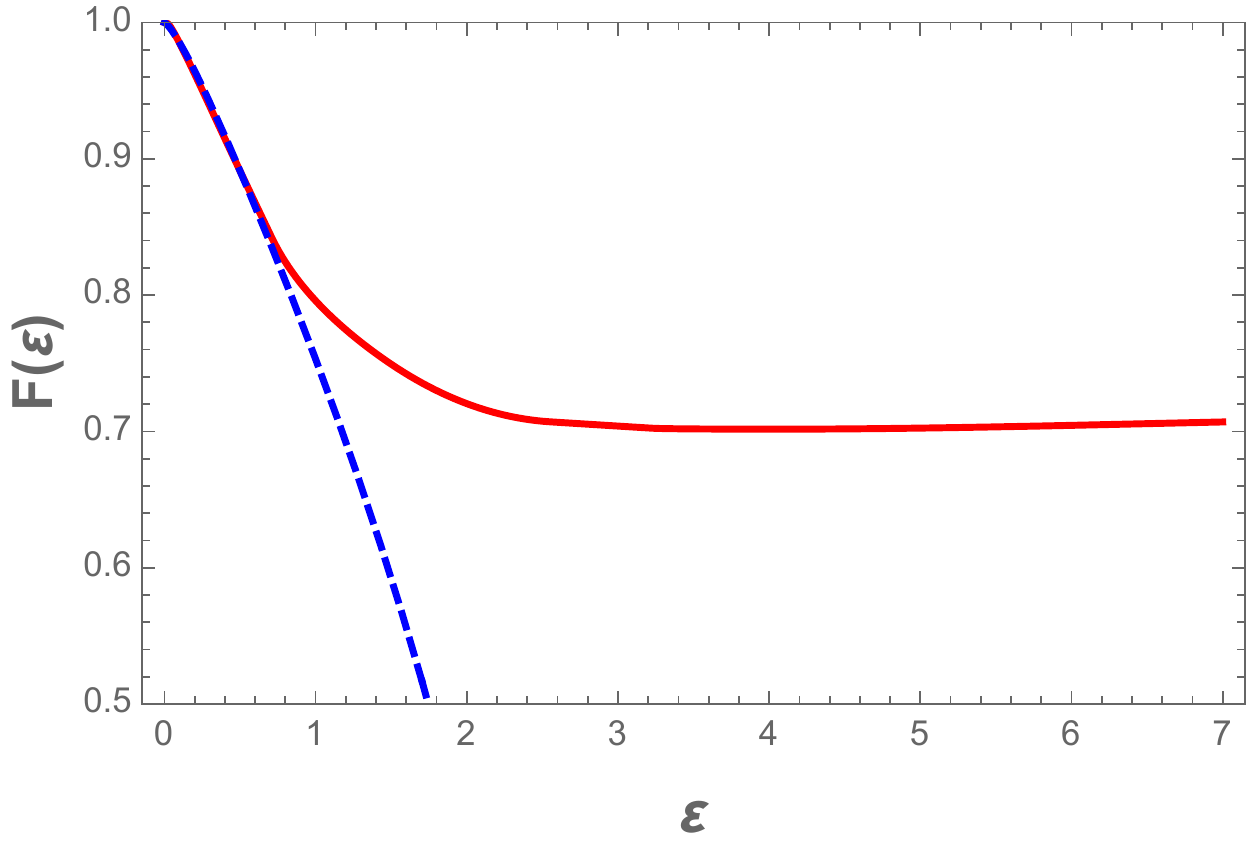}
\\
\end{tabular}
\end{center}
\vskip-.4cm
\caption{
The instanton suppression holy grail function $F$ \eqref{eq:Fspfin}(shown in red) 
plotted over a broad energy range, $0\le \varepsilon=\frac{E}{\pi m_W/\alpha_w}\le 30$. 
The horizontal dotted line indicates the minimal value of $F\simeq0.70$ that occurs at $\varepsilon=4$. 
The matching between the high-energy and the low-energy regimes is smooth as shown on the plot on the right. The known analytic expression 
\eqref{eq:Fsp2} for $F(\varepsilon)$ (dashed blue line) shows a good match with our result for $\varepsilon <1$. }
\label{fig:F}
\end{figure}

The instanton suppression factor $F_*(\varepsilon)$ is plotted in Fig.~\ref{fig:F} over the entire energy range. It can be seen from the data in the 
Table~\ref{Tab:one} and from Figure~\ref{fig:F} that instanton--anti-instanton separations are never below $\chi_{\rm min} \simeq 2.06$, thus staying clear from the perturbative region. It should be noted from the data in Table~\ref{Tab:one} that our instantons remain small over the whole range of energies considered, justifying the use of our approximations and the $I\bar{I}$ action is large, $S\ge 0.867$. The instanton suppression factor 
is significant with the minimal value of the holy-grail function,  $F\simeq 0.70$, loosing only 30\% of the original 't Hooft instanton suppression. At energies lower and higher than the critical energy, $\varepsilon=\frac{E}{\pi m_W/\alpha_w}=4$, the suppression of the instanton rate increases further. Electroweak instantons remain exponentially suppressed by at least $e^{-\frac{4\pi}{\alpha_w} 0.70}$ and are unobservable in high energy 2 particle collisions for any value of the energy.

In Table~\ref{Tab:two}, we present the saddle-point values of $\hat{\rho},\chi$ and $F$ which come from a numerical solution of the full saddle-point equations coming from Eq.~\eqref{eq:fullF}, taking into account all known terms, i.e. not neglecting the higher order $\hat{\rho}$ terms as we did above. It can be seen from this data that the inclusion of these higher order terms does not significantly affect our conclusions; the holy-grail function, $F$, still does not drop far below 0.7, leading to an exponential suppression of electroweak instanton processes, indicating that they should be unobservable at colliders. 

\begin{table}[]
\begin{center}
\begin{tabular}{|r||r|r|r||r|}
\hline
$\varepsilon$ & $\chi\,\,$ & $\hat{\rho}\,\,\,$ & $S(\chi)$ & $F_*\,\,$\\
\hline\hline
30   &   2.143 & 0.018 & 0.879 & $0.735$\\ 
20   &   2.187 & 0.029 & 0.885 & $0.727$\\ 
10   &   2.102 & 0.061& 0.873 & $0.713$\\ 
7.0   &   2.076 & 0.090 & 0.869 & $0.706$\\
5.0   &   2.065 & 0.130 & 0.867 & $0.700$\\ 
\hline
4.0   &   2.056 & 0.165 & 0.866 & $0.697$\\ 
\hline
3.0   &   2.065 & 0.225& 0.867 & $0.696$\\ 
2.0   &   2.100 & 0.352& 0.873 & $0.699$\\ 
0.5   &   2.632 & 0.394& 0.931 & $0.861$\\ 
\hline
\end{tabular}
\end{center}
\vskip-.4cm
\caption{
Numerical solutions of the saddle point equations coming from Eq.~\eqref{eq:fullF} in the same energy range as in Table 1. Instantons remain exponentially suppressed at all energies with the minimal value of $F_*$ at the critical energy $\varepsilon\simeq 3$.}
\label{Tab:two}
\end{table}

\section{More on hard quantum corrections}

In this section we will give a brief overview of quantum corrections arising from interactions between gauge bosons \cite{Mueller:1990ed} and between fermions in the initial state. In our calculation in the preceding section these quantum effects have been accounted for by the $\frac{\alpha_w}{16\pi}(\rho^2+\bar{\rho}^2)E^2\log \left({E^2}/{m_W^2}\right)$ term in the holy grail function. 

\subsection{Quantum corrections from vector bosons}
\label{sec:QC}

\begin{figure}[]
\begin{center}
\hspace{-.6cm}
\includegraphics[width=0.8\textwidth]{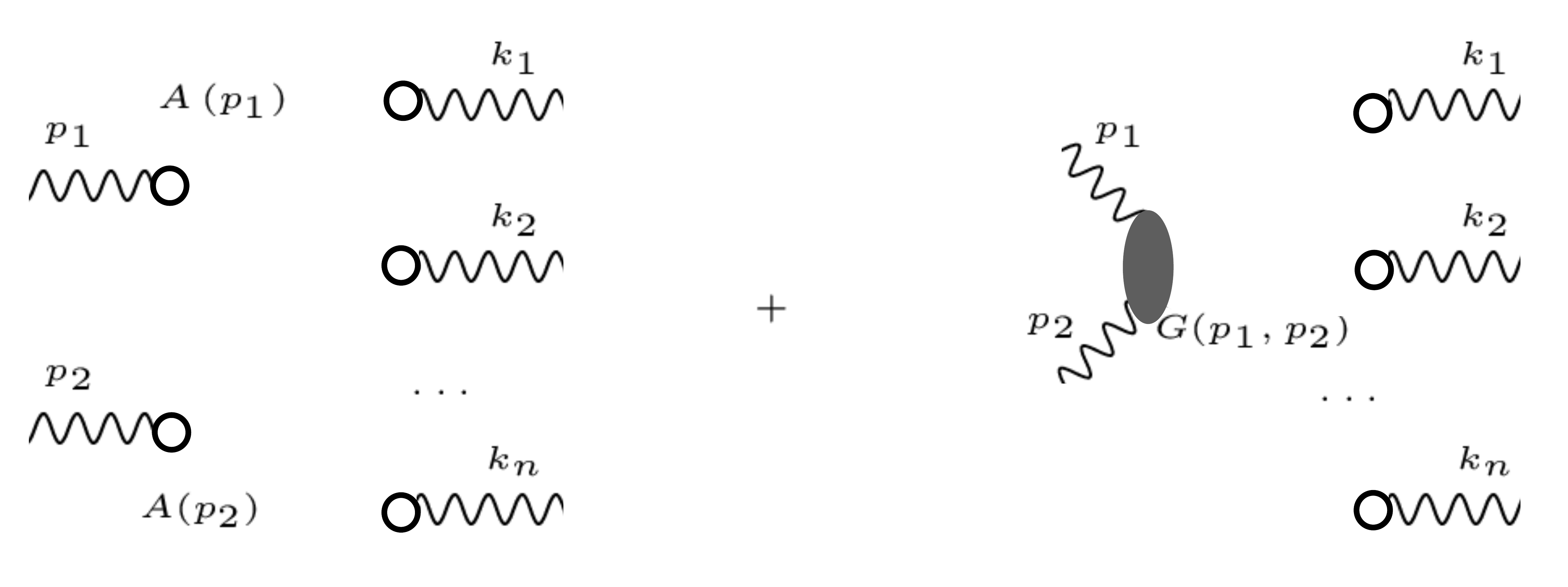}
\end{center}
\vskip-.6cm
\caption{
The classical contribution and the leading-order correction to the $2\to n$ gauge-boson amplitude in the instanton background. Each tadpole represents an insertion of the instanton gauge field into the path integral, and the shaded blob corresponds to the vector-boson propagator  $G^{ab}_{\mu\nu}\left(p_{1},p_{2}\right)$ in the instanton background.}
\label{fig:PT1}
\end{figure}

We first consider the instanton-generated $2\to n$ process in a pure gauge theory. The classical result and the leading order correction in the instanton perturbation theory to this amplitude are shown in Fig.~\ref{fig:PT1}. We are concentrating here on quantum corrections due to initial state particles; other interactions involving initial-final and final-final state interactions are already accounted for in the optical theorem approach.

In order to capture the high-energy behaviour of the perturbative expansion around the instanton, we need the expression for the gauge-field propagator in the instanton background in the high-energy limit. The required result was derived by Mueller in  \cite{Mueller:1990qa}.
Starting from the full expression for the propagator from \cite{Brown:1977eb}, upon continuation to Minkowski space and taking the on-shell limit, $p_1^2=0=p_2^2$, along with the high-energy limit, $2 p_1 p_2 =s \gg 1/\rho^2$, where $\rho$ is the instanton size, the result of \cite{Mueller:1990qa} is,
\begin{equation}
G^{ab}_{\mu\nu}\left(p_{1},p_{2}\right)\, \to\,-\frac{g^{2}\rho^{2}s}{64\pi^{2}} \log \left(s\right)A_{\mu}^{a}\left(p_{1}\right)A_{\nu}^{b}\left(p_{2}\right),
\label{eq:Mproplim}
\end{equation}
where $A_{\mu}^{a}\left(p_{1}\right)$ and $A_{\nu}^{b}\left(p_{2}\right)$ are the instanton solutions for the gauge fields in momentum space. The key point of this expression is that in the high-energy limit the instanton vector propagator is proportional to the product of the classical instanton fields -- this fact will be relevant for the resumation of these effects. The coefficient in front of the instanton fields involves the large quantity $\rho^2 s\log s$ which compensates for the smallness of the perturbative coupling $g^2$.

The two initial-state vector bosons are represented in the instanton perturbation theory as the product of two instanton field configurations 
$A_{\mu}^{a}\left(p_{1}\right) A_{\nu}^{b}\left(p_{2}\right)$, shown as the two tadpoles on the left in the first diagram in  Fig.~\ref{fig:PT1}. The first quantum correction to this initial state comes from the propagator $G^{ab}_{\mu\nu}\left(p_{1},p_{2}\right)$, as shown in the second diagram in Figure~\ref{fig:PT1}. The combined result of these two diagrams is the insertion of the factor, 
\begin{equation}
\left(1-\frac{g^{2}\rho^{2}s}{64\pi^{2}}\log s \right) A_{\mu}^{a}\left(p_{1}\right)A_{\nu}^{b}\left(p_{2}\right),
\label{eq:Mprop2}
\end{equation}
into the path integral to represent the 2-particle initial state in the corresponding $(2+n)$-point correlator. 

In Ref.~\cite{Mueller:1990ed} Mueller computes the higher-order corrections to this result 
by summing over all loop-level perturbative diagrams to order $N$ involving the two initial-state vector bosons in the instanton background. The result is,
\begin{equation}
\sum_{r=1}^{N} \frac{1}{r!}\left(-\frac{g^{2}\rho^{2}s}{64\pi^{2}}\log \left(s\right)\right)^r \,A_{\mu}^{a}\left(p_{1}\right)A_{\nu}^{b}\left(p_{2}\right).
\label{eq:Mprop3}
\end{equation}
The equation is justified in the limit $\rho^2 s\to \infty$ with $g^2 \to 0$ such that any power of $g^2 \rho^2 s$ is counted as of order 1.

In the limit $N\to \infty$ we obtain the exponential factor,
\begin{equation}
\exp\left[-\, \frac{\alpha_w}{16\pi}\rho^2E^2\, \log \left({E^2}/{m_W^2}\right) \right]\,.
\label{eq:Mprop4}
\end{equation}
The contribution of the Mueller term above to the cross-section is then,
\begin{equation}
\exp\left[-\, \frac{\alpha_w}{8\pi}\rho^2E^2\, \log \left({E^2}/{m_W^2}\right) \right] \,=\, 
\exp\left[-\,\frac{4\pi}{\alpha_w}\, \frac{\hat{\rho}^2\varepsilon^2}{16} \,\log({\varepsilon\, \pi}/{\alpha_w})
\right]\,,
\label{eq:Mprop4sig}
\end{equation}
in agreement with the third term on the right hand side of the expression \eqref{eq:calF22} that we used in the previous section for holy grail function $F$.

\medskip

Can one formally justify taking the $N\to \infty$ limit in the expression \eqref{eq:Mprop3}?
To be able to do this, we have to demonstrate that the sum in \eqref{eq:Mprop3} correctly approximates instanton perturbation theory 
for values of $N$ greater than the critical value 
$N_{\rm crit} \sim \frac{\alpha_w}{16\pi}\rho^2 s$. (This critical value is given by the argument of the exponent, and for values of $r$ much above 
this $N_{\rm crit}$ each term term in the sum \eqref{eq:Mprop3} is parametrically smaller than the exponent \eqref{eq:Mprop4}, so it can be legitimately 
dropped.)
The expression \eqref{eq:Mprop3} was derived in~\cite{Mueller:1990ed} by retaining only the most dominant terms in the high-energy limit $s\rho^2 \gg 1$.
The $N^{\rm th}$ term in \eqref{eq:Mprop3} follows from the $(N-1)$-loop level diagrams shown in Fig. 9 in Ref.~\cite{Mueller:1990ed}. To justify the approximation where all sub-leading terms are not retained, it is required that $\rho^2 s/N^{2} \gg 1$. Hence we can trust the sum in  \eqref{eq:Mprop3}  
only up to $N< \sqrt{\rho^2 s}$. This implies the window for $N$ in the form~\cite{Mueller:1990ed},
\begin{equation}
\frac{\alpha_w}{16\pi}\rho^2 s \,<\, N\,<\, \sqrt{\rho^2 s}\,.
\label{eq:ineqM}
\end{equation}
Mueller discusses this limit in the low-energy regime where the function in the exponent in \eqref{eq:Mprop4}  is sub-leading relative to the other terms in the holy grail function (specifically, it goes as $\varepsilon^{8/3}$ relative to the leading contributions $\varepsilon^{4/3}$ where $\varepsilon \ll 1$).
We want to consider instead the high-energy regime $\varepsilon \gtrsim 1$ where the Mueller quantum effect  \eqref{eq:Mprop4} plays the dominant role in cutting off the instanton size.
This implies that the magnitude of the Mueller term is comparable to the other terms in the exponent (e.g. the constant term in the instanton--anti-instanton action) so $ \frac{\alpha_w}{16\pi}\rho^2 s \sim \frac{2\pi}{\alpha_w} \, {C}$ where $C$ is of the order of $1-F_*$, roughly 0.3 around the critical energy (and less than that otherwise). This gives the characteristic value of $\rho^2 s \sim \frac{32\pi^2}{\alpha_w^2} \, { C}$ where the Mueller exponent becomes important. Plugging this into the inequality \eqref{eq:ineqM}, we find,
\begin{equation}
\frac{2\pi}{\alpha_w} \,C \,<\, N\,<\, \frac{4\sqrt{2}\pi}{\alpha_w} \,\sqrt{C}\,.
\label{eq:ineqV}
\end{equation}
Sine we expect $C< 1$, this `window of opportunity' for $N$ is certainly not empty.  So, while there is no rigorous proof of exponentiation, we are comfortably optimistic that the Mueller correction does exponentiate and \eqref{eq:Mprop4} holds.\footnote{There are of course many known examples of resummed perturbation theory that give rise to decaying exponentials. These include Sudakov form-factors~\cite{Collins:1989bt} and further examples are provided by resummed leading-loop corrections to tree-level $1\to n$ perturbative amplitudes near mass thresholds \cite{Libanov:1994ug} contributing to another incarnation of a perturbative holy grail function. }

We should point out, however, that we cannot a priori exclude higher-order perturbative corrections to the Mueller term in the exponent of \eqref{eq:Mprop4},
that would be of the form, 
\begin{equation}
\alpha_w\, (\alpha_w \rho^2 E^2)^2 \, \sim\,  \frac{1}{\alpha_w} (\hat{\rho}\varepsilon)^4\,.
\label{eq:Mprop4H}
\end{equation}
Such term is subdominant relative to the Mueller term $\sim \alpha_w \rho^2 E^2$  in the limit $\alpha_w\to 0$, $\rho^2 E^2 \to \infty$ with $\alpha_w \rho^2 E^2$ held fixed, and as such can potentially be generated in instanton perturbation theory. Accounting for this correction, the exponent in \eqref{eq:Mprop4sig} would be modified as follows,
\begin{equation}
\exp\left[-\,\frac{4\pi}{\alpha_w} \left( \frac{\hat{\rho}^2\varepsilon^2}{16} \,+A\, \hat{\rho}^4\varepsilon^4 \right)\log({\varepsilon\, \pi}/{\alpha_w})
\right]\,,
\label{eq:Mprop4sig2}
\end{equation}
where $A$ is a constant.
Parametrically, the saddle-point solution for $\hat{\rho}$ in the high-energy regime goes as  $\hat{\rho} \sim 1/\varepsilon$, according to \eqref{eq:solrhoV}, and both terms are of the same order. This implies that the higher-order corrections to Mueller term in the exponent (if present) could modify our conclusions. However, for a positive constant $A$ the instanton rates would become only more suppressed. On the other hand, for a negative $A$ we one can always reach a sufficiently high energy $\varepsilon_\star$ where the instanton size $\rho$ is no longer cut-off neither by Mueller term, not by the Higgs vev, and the IR problem of large instantons is re-introduced, with the integral over $\rho$ now {\it exponentially} divergent. We find such a UV/IR mixing\footnote{This refers to the fact that at all energies below 
$\varepsilon_\star$ instanton cross-sections are IR safe, while at energies above $\varepsilon_\star$, contributions of large instantons become exponentially divergent.}
in the electroweak theory to be highly unlikely, and will assume on physical gorunds that no exponentially growing corrections to the Mueller term are present.

\medskip

 Up to now we have only considered the initial-state corrections arising from interactions between gauge bosons, but as we will show in the next sub-section,
the same result \eqref{eq:Mprop4} applies equally well to fermions in the initial state, which is our main case of interest for the process \eqref{eq:inst1}.
The reason is that the initial state fermions will radiate gauge bosons. 
 
 \newpage

\subsection{Quantum corrections from initial fermions}

\begin{figure}[]
\begin{center}
\hspace{-.6cm}
\includegraphics[width=0.82\textwidth]{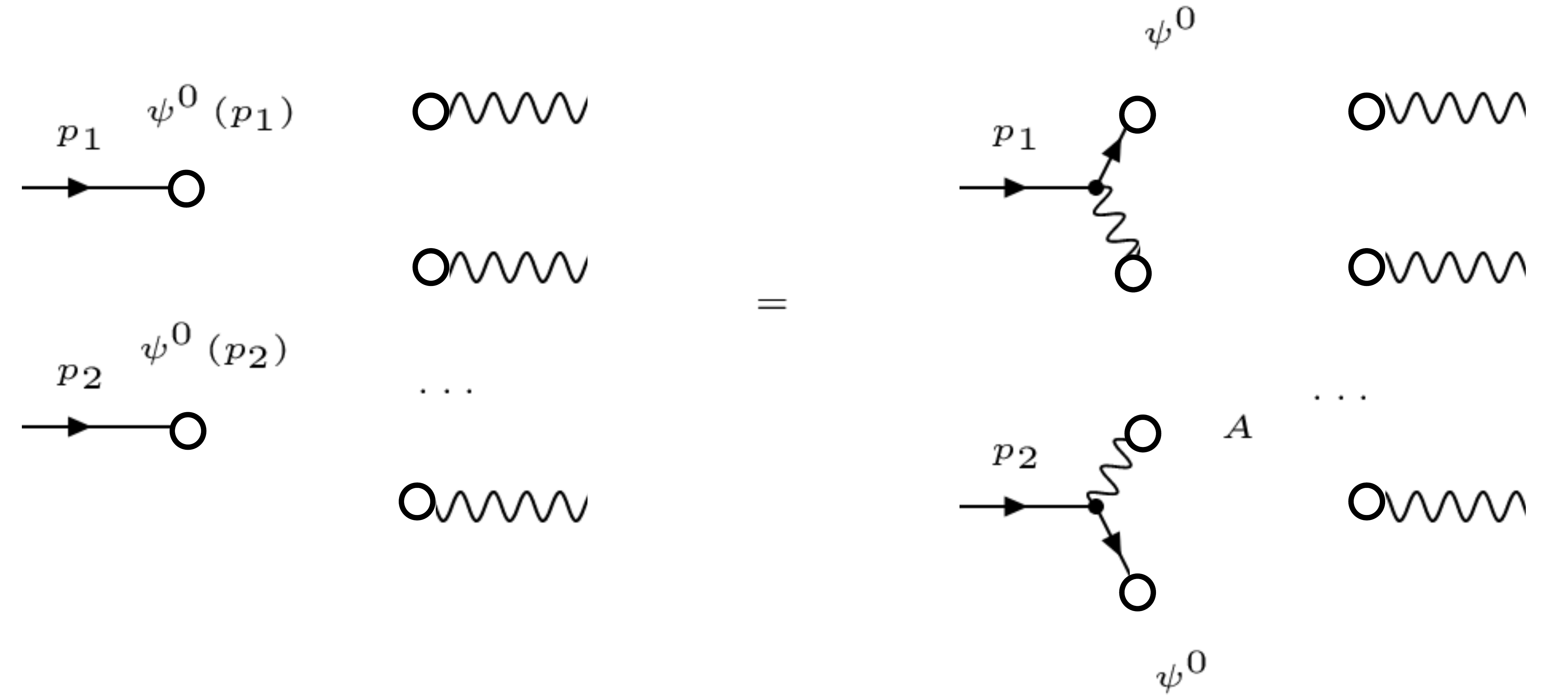}
\end{center}
\vskip-.6cm
\caption{
The leading-order diagram for the instanton process \eqref{eq:inst12}  and its equivalent representation using the Dirac equation \eqref{eq:DirI}.}
\label{fig:PT3}
\end{figure}

We will now consider the instanton-generated process with the fermions in the initial state. For simplicity consider the 2 fermions into $n$ vector bosons
process in a theory with 1 fermion flavour (generalisation to the full process \eqref{eq:inst1} in the Standard Model is trivial as the differences involve only the final state),
\begin{equation}
 q\, +\, q \, \to\,  n \,W \,.
 \label{eq:inst12}
 \end{equation} 
 At the leading order in instanton perturbation theory, the two quarks $q$ are represented by insertions of the instanton fermion zero modes $\psi^{(0)}$, and the $n$ final-state vector bosons $W$ are represented by insertions of $n$ instanton gauge-field configurations $A_\mu$ into the corresponding correlator. This leading-order classical contribution to the process \eqref{eq:inst12} is represented by the first diagram in Fig.~\ref{fig:PT3}.
 
\begin{figure}[h]
\begin{center}
\hspace{-.6cm}
\includegraphics[width=0.5\textwidth]{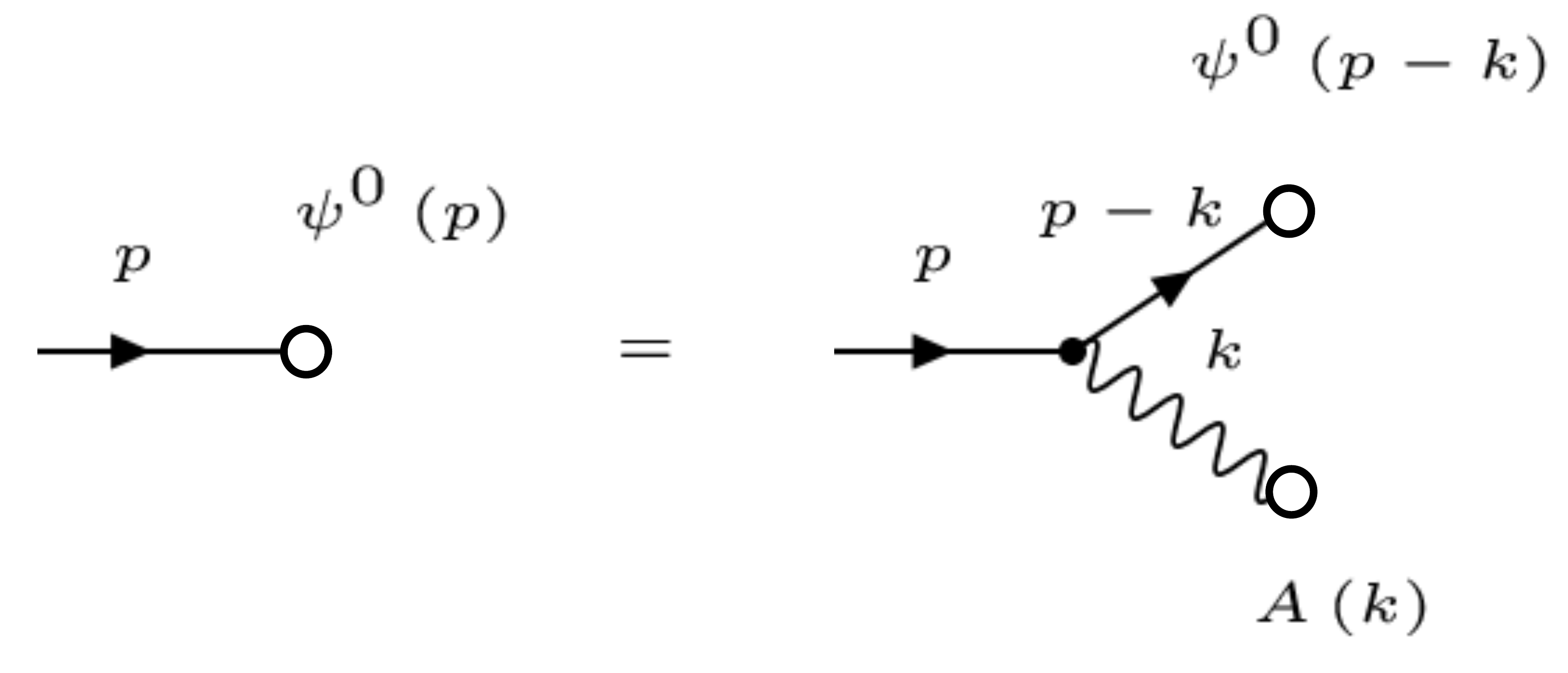}
\end{center}
\vskip-.9cm
\caption{
Pictorial representation of the Dirac equation \eqref{eq:DirI}.}
\label{fig:PT2}
\end{figure}

The fermion zero modes $\psi^{(0)}(x)$ are the non-trivial (and normalisable) solutions of the Dirac equation in the instanton background $A_\mu(x)$,
\begin{equation}
\gamma^\mu (\partial_\mu - i g A_\mu) \psi^{(0)}(x) \,=\, 0\, .
\label{eq:DirI}
\end{equation}
This implies that we can represent the LSZ-reduced fermion zero mode in terms of the fermion-vector boson vertex, as indicated in Fig.~\ref{fig:PT2}.

\begin{figure}[]
\begin{center}
\hspace{-.4cm}
\includegraphics[width=0.82\textwidth]{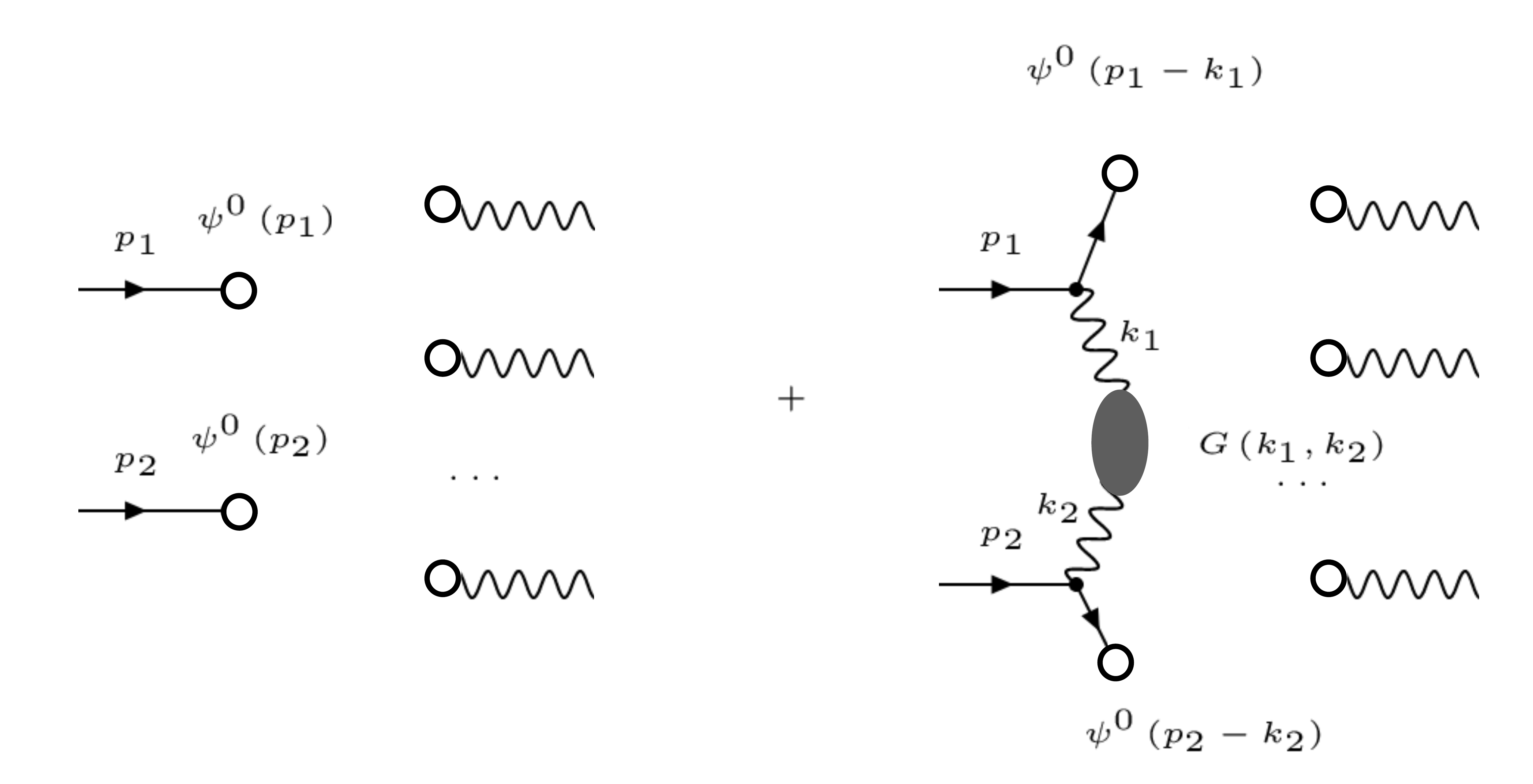}
\end{center}
\vskip-.6cm
\caption{
The leading and the next-to-leading order contributions to the instanton process with the fermions in the initial state.}
\label{fig:PT4}
\end{figure}

Using this we have the equivalent representation of the leading-order contribution to the process \eqref{eq:inst12} shown as the diagram on the right in Fig.~\ref{fig:PT3}.
At the next-to-leading order in the instanton perturbation theory we simply connect the two gauge fields in this diagram by the propagator in the instanton background. This correction is represented by the second diagram in Fig.~\ref{fig:PT4}.

Using the known high-energy limit of the instanton vector propagator, we can write down the combined contribution of the two diagrams in  Fig.~\ref{fig:PT4} as,
\begin{equation}
\left(1-\frac{g^{2}\rho^{2}s}{64\pi^{2}}\log s \right)  \psi^{(0)}(p_{1})  \psi^{(0)}(p_{2}),
\label{eq:Mprop2ff}
\end{equation}
in a complete analogy with the corresponding expression for the initial-state gauge fields in \eqref{eq:Mprop2}.
Now the problem has been reduced to keeping the dominant effects in the $\rho^2 s\to \infty$ limit in the instanton perturbation theory in the gauge sector.
The outcome of this procedure is, as in the previous sub-section, the exponential factor in \eqref{eq:Mprop4},
\begin{equation}
\exp\left[-\, \frac{\alpha_w}{16\pi}\rho^2E^2\, \log \left({E^2}/{m_W^2}\right) \right] \, \psi^{(0)}(p_{1})  \psi^{(0)}(p_{2}),
\label{eq:Mprop4ff}
\end{equation}
which reproduces the instanton contribution to the final term of the holy grail function $F$ that we used in \eqref{eq:calF1}.

\bigskip
\section{\label{Sec:Conclusion}Conclusions}

In this paper we have presented an updated calculation for the electroweak instanton cross section, combining the optical theorem formalism
with the inclusion of quantum corrections arising from the initial state interactions, which result in the dominant effect in the high energy limit. 
A notable feature of our approach is that we were able to justify it in the high energy limit, where the energy is much greater than the sphaleron
mass and where many earlier instanton-based estimates had failed. This was based in part on an assumption that there are no additional exponentially growing (and unphysical, as explained at the end of section~\ref{sec:QC}) contributions generated at higher orders in instanton perturbation theory.

By interpolating between the high-energy and the low-energy results we have shown that electroweak instanton-dominated $2\to n$ processes remain exponentially suppressed at all energies and would not be observable, even in principle, at any collider. We did not provide a full calculation of the cross section involving all pre-exponential factors (which is possible, e.g, following the methods of \cite{Khoze:2020tpp}) as, given the magnitude of the exponential suppression, the cross section would still not reach an observable level. 

\medskip

It is interesting to compare our results to those of Refs.~\cite{Bezrukov:2003er,Bezrukov:2003qm}. The authors numerically computed rates for the electroweak 
transitions across the sphaleron barrier starting with processes containing a large number of particles $\sim N/\alpha_w\gg 1$ in the initial state, before taking the limit $N\to 0$ and assuming that it is smooth.  The authors of~\cite{Bezrukov:2003er,Bezrukov:2003qm}
 also obtained that the cross section was exponentially suppressed at all energies. Their numerical results do not agree precisely with ours, but this should not be expected given the different starting points, methodologies and approximations. Nevertheless we consider the two approaches to be complimentary and they both lead to the same  conclusions.

\medskip

In our calculation, the electroweak instantons did not acquire a very substantial compensation of the
original 't Hooft suppression factor $e^{-4\pi/\alpha_w}$ that would correspond to $F=1$. Our results in Tables~\ref{Tab:one} and \ref{Tab:two}  (see also Fig.~\ref{fig:F}) for the instanton holy grail function $F$ show that the 
 the 't Hooft suppression in the exponent is reduced by not more than  $\sim 30\%$ over the entire energy range.
This justifies neglecting contributions from the multi-instanton configurations
that were considered in the `premature unitarization' approach in Refs.~\cite{Zakharov:1990xt,Maggiore:1991vi,Veneziano:1992rp}
and were argued to suppress the instanton result if the function $F$ was reduced by $\gtrsim 50\%$ in the original instanton calculation.

\section*{Acknowledgements}

We thank Frank Krauss, Matthias Schott and Michael Spannowsky helpful discussions.
This work was supported in part by the STFC grant ST/P001246/1.

\newpage

\bibliographystyle{inspire}
\bibliography{main}

\clearpage
\appendix
\include{variables}

\end{document}